\documentclass[journal]{IEEEtran}
\usepackage{cite}

\ifCLASSINFOpdf
\usepackage[pdftex]{graphicx}
\DeclareGraphicsExtensions{.pdf,.jpeg,.png}
\else
\usepackage[dvips]{graphicx}
\DeclareGraphicsExtensions{.eps}
\fi
\usepackage{epsfig,epsf,epstopdf,graphicx}
\usepackage[cmex10]{amsmath}
\usepackage{amssymb}
\usepackage{amsthm }
\usepackage{nicefrac}
\usepackage{hyperref}
\usepackage{xcolor}
\hypersetup{colorlinks=true}
\usepackage{tabularx}

\theoremstyle{theorem}

\theoremstyle{definition}

\theoremstyle{plain}

\theoremstyle{plain}

\newcommand{\onevec}{{\bf{1}}}

\newcommand{\xb}{{\textbf{x}}}
\newcommand{\yb}{{\textbf{y}}}

\newcommand{\Db}{{\textbf{D}}}

\newcommand{\nb}{{\textbf{n}}}

\newcommand{\Ib}{{\textbf{I}}}

\newcommand{\rb}{{\textbf{r}}}

\newcommand{\thetab}{{\mbox{\boldmath $\theta$}}}

\newcommand{\Lb}{\mathbf{L}}

\newcommand{\Wb}{\mathbf{W}}

\usepackage{algorithm}
\usepackage{multirow}
\usepackage{algcompatible}
\usepackage[bottom]{footmisc}

\PassOptionsToPackage{dvipsnames}{xcolor}
\usepackage{tikz}
\usetikzlibrary{calc}

\ifCLASSOPTIONcompsoc
  \usepackage[caption=false,font=normalsize,labelfont=sf,textfont=sf]{subfig}
\else
  \usepackage[caption=false,font=footnotesize]{subfig}
\fi
\usepackage{fixltx2e}
\usepackage{afterpage}
\usepackage{float}
\usepackage{stfloats}

\makeatletter

\begin{document}

%
\title{Bayesian Convolutional Neural Networks for Prior Learning in Graph Signal Recovery}

\author{Razieh~Torkamani,~Arash~Amini,~Hadi~Zayyani,~and~Mehdi~Korki
\thanks{This work was supported by the Iran National Science Foundation (INSF) (grant number
4012697) (Corresponding author: R.~Torkamani).}
\thanks{R.~Torkamani is with Department
of Electrical Engineering, Faculty of Engineering, Bu-Ali Sina University, Hamedan, Iran (e-mail: r.torkamani@basu.ac.ir).}
\thanks{A.~Amini is with Department
of Electrical Engineering, Sharif University of Technology, Tehran, Iran (e-mail: aamini@sharif.ir).}
\thanks{H.~Zayyani is with Department
of Electrical and Computer Engineering, Qom University of Technology (QUT), Qom, Iran (e-mail: zayyani@qut.ac.ir).}
\thanks{M.~Korki is with School of Engineering, Swinburne University of Technology, Melbourne, Australia (e-mail: mkorki@swin.edu.au).}
\vspace{-0.5cm}}


\maketitle
\thispagestyle{plain}
\pagestyle{plain}

\begin{abstract}
Graph signal recovery (GSR) is a fundamental problem in graph signal processing, where the goal is to reconstruct a complete signal defined over a graph from a subset of noisy or missing observations. A central challenge in GSR is that the underlying statistical model of the graph signal is often unknown or too complex to specify analytically. To address this, we propose a flexible, data-driven framework that learns the signal prior directly from training samples. We develop a Bayesian convolutional neural network (BCNN) architecture that models the prior distribution of graph signals using graph-aware filters based on Chebyshev polynomials. By interpreting the hidden layers of the CNN as Gibbs distributions and employing Gaussian mixture model (GMM) nonlinearities, we obtain a closed-form and expressive prior. This prior is integrated into a variational Bayesian (VB) inference framework to estimate the posterior distribution of the signal and noise precision. Extensive experiments on synthetic and real-world graph datasets demonstrate that the proposed BCNN-GSR algorithm achieves accurate and robust recovery across a variety of signal distributions. The method generalizes well to complex, non-Gaussian signal models and remains computationally efficient, making it suitable for practical large-scale graph recovery tasks.

\end{abstract}


%
\IEEEpeerreviewmaketitle

\section{Introduction}
\label{sec:Intro}
In a wide range of applications --including sensor networks \cite{Shum13}, biological networks, machine learning, data science \cite{Ortega18}, and image processing \cite{Ortega18} the signals of interest are naturally defined over graphs. In many of these scenarios, the data does not reside in a conventional metric space but is instead situated on irregular or non-Euclidean domains. As a result, traditional signal processing techniques  often fall short for effectively analyzing or processing such data. To address this challenge, Graph Signal Processing (GSP) has emerged as a powerful framework for modeling signals supported on graphs and for capturing complex interactions among them. The core objective of GSP is to generalize classical signal processing tools to graph-structured data, enabling processing methods that are sensitive to both the topology of the graph and the characteristics of the graph signals themselves \cite{Shum13}-\cite{Torka22}.

GSP has gained considerable attention in recent literature \cite{Sand13}- \cite{Tork21} due to its ability to model and analyze data residing on irregular domains. Many GSP algorithms are grounded in the spectral processing of graph signals, typically leveraging the Graph Fourier Transform (GFT) \cite{Ortega18}, \cite{Pes08}-\cite{Pes10}. Two primary approaches exist for defining the GFT: one maps graph signals onto the eigenvectors of the graph Laplacian matrix \cite{Pes08}, and the other uses the eigenvectors of the graph adjacency matrix \cite{Ortega18}. For undirected graphs, the GFT is traditionally defined using the Laplacian eigenvectors, which supports spectral clustering and minimizes the variation of the signal’s $\ell_2$-norm over the graph. For directed graphs, an alternative GFT is often used, where the adjacency matrix serves as the graph shift operator \cite{Pus08}, \cite{Pus081}. Another approach in \cite{Sard16} proposes that the GFT basis should be formed from orthogonal functions that minimize the cut size in directed graphs. Further developments, such as in \cite{Cheng23} define GFTs using singular value decompositions (SVDs) of Laplacians on Cartesian product graphs formed by two directed graphs.

A key problem in GSP is Graph Signal Recovery (GSR), which involves reconstructing a complete graph signal from a subset of noisy or incomplete observations.
For example, the method proposed in \cite{Chen15} recovers smooth signals through variation minimization using the Alternating Direction Method of Multipliers (ADMM). The work in \cite{Java24} introduces an algorithm for simultaneous inference of graph topology and signal from incomplete time-series data, using the Block Successive Upper Bound Minimization (BSUM) framework. In \cite{Khon23}, a multilayer GSR algorithm is developed for signals defined over multilayer graphs, incorporating robustness to outliers via median truncation in gradient descent. Additionally, \cite{Jiang20} proposes two distributed GSR algorithms using $\ell_1$ and $\ell_2$-norm-based optimization, which rely on data exchanges among localized neighborhood subgraphs. Adaptive signal recovery approaches have also been explored. Methods such as Least Mean Squares (LMS) \cite{Loren16}, Normalized LMS (NLMS) \cite{Spel20}, Proportionate LMS \cite{Tork23}, and Proportionate Recursive Least Squares (RLS) \cite{Naem24} have been adapted to graph domains for recovering signals from partial and noisy measurements in an online manner. Existing approaches to GSR often assume that the underlying signal follows a known statistical model, such as a Gaussian Markov Random Field (GMRF) \cite{Tork21}, or rely on smoothness priors encoded by the graph Laplacian. However, these assumptions may not hold in practice, as the true distribution of the graph signal is often unknown or too complex to model analytically.

To overcome this limitation, data-driven methods have been proposed to learn graph signal models directly from data. Neural networks, especially Graph Neural Networks (GNNs), have shown promise for graph signal sampling and recovery tasks \cite{Chen20}.
In \cite{Chen20}, GNNs estimate mutual information between neighboring nodes, and recovery is performed using an unrolled optimization algorithm. Similarly, \cite{Gama19}, proposes CNN-inspired architectures for GSP, and \cite{Gao21} develops a stochastic gradient descent-based training framework for GNNs in random network settings. However, conventional deep neural networks (DNNs) and GNNs typically produce point estimates and lack mechanisms to quantify uncertainty in their predictions. Moreover, they often assume fixed model structures, which can limit their flexibility when the true data distribution deviates from prior assumptions.

In this paper, we propose a Bayesian Convolutional Neural Network for Graph Signal Recovery (BCNN-GSR). Our method combines the expressiveness of deep learning with the rigor of Bayesian inference (by treating weights and outputs as random variables and computing their marginal distributions) to build a robust, uncertainty-aware recovery framework. The core idea is to use a Bayesian CNN to learn the prior distribution of the graph signal from training data, and then employ variational Bayesian (VB) inference to recover the signal from noisy observations. Unlike conventional CNNs, which operate in the Euclidean domain, our proposed CNN integrates graph convolutional filters based on Chebyshev polynomials to explicitly exploit the underlying graph topology.

\begin{figure}[!t]
\centering
{\includegraphics[width=.95\linewidth]{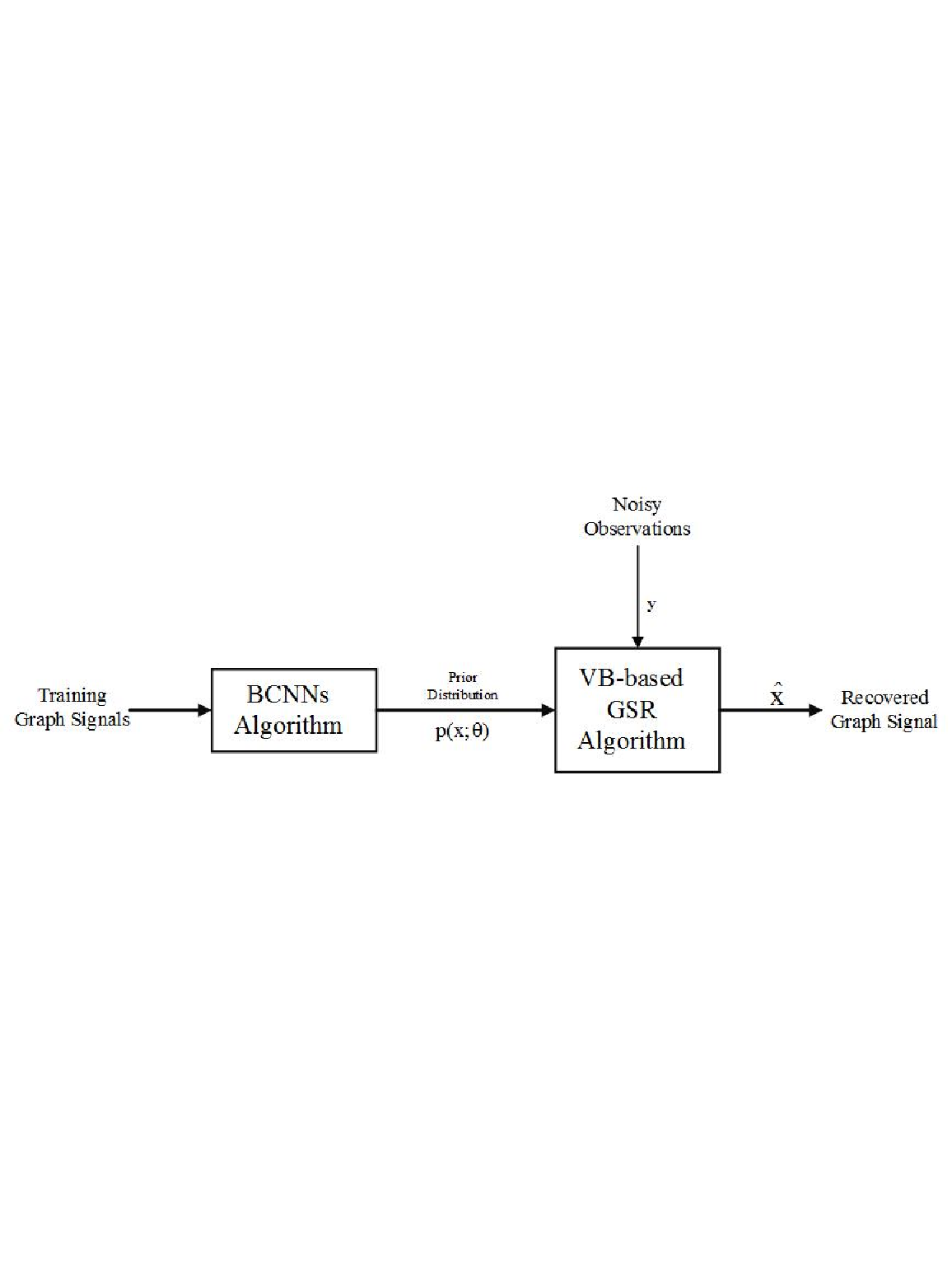}}%
\caption{Block diagram of the proposed algorithm for graph signal recovery.}
\label{fig1}
\end{figure}

In this work, we model the hidden layers of DNNs as Gibbs distributions, effectively formulating a hierarchical Bayesian model for graph signal priors.
To efficiently incorporate graph structure into learning-based models, we adopt Chebyshev polynomial filters \cite{Deff16} as the core component of our graph convolution operation. These filters approximate spectral graph convolutions by expressing them as linear combinations of Chebyshev polynomials applied to a rescaled graph Laplacian. This formulation eliminates the need for explicit eigendecomposition, substantially reducing computational complexity and enabling scalability to large graphs. Furthermore, the use of $K$-order Chebyshev polynomials ensures that each filter operates within a node’s $K$-hop neighborhood, preserving the locality of the graph structure. By enabling localized, efficient, and topology-aware filtering in the spectral domain, Chebyshev polynomial filters provide a powerful mechanism for learning meaningful representations of graph signals, particularly valuable for GSR tasks.
We then use this learned prior in a VB inference algorithm to estimate posterior distributions and perform GSR. The overall architecture of the proposed BCNN-GSR method is illustrated in Fig. \ref{fig1}.

The main contributions of this paper are as follows:
\begin{itemize}
\item We introduce, for the first time, a Bayesian deep learning framework for graph signal recovery by modeling the hidden layers of a convolutional neural network as Gibbs distributions. This probabilistic interpretation enables the learning of expressive, data-driven priors for graph signals.
\item We incorporate Chebyshev polynomial filters into the CNN architecture to perform localized, efficient graph convolution. This ensures that the learned filters respect the underlying graph topology without requiring expensive eigendecomposition, making the method scalable to large graphs.
\item We develop a variational Bayesian (VB) inference procedure that leverages the learned prior distribution to estimate both the graph signal and the noise precision. This results in a robust posterior that adapts to complex, non-Gaussian signal distributions.
\end{itemize}

We validate the effectiveness of our approach through experiments on both synthetic and real-world graph signals, including temperature sensor data. Results demonstrate that BCNN-GSR consistently outperforms state-of-the-art methods such as VB-GSR \cite{Tork21}, GNN-GSR \cite{Chen20}, and Joint Inference of the network topology and Signals over Graphs (JISG) \cite{Ioan19}, particularly when the true signal distribution deviates from the Gaussian prior assumed in other models.

The rest of the paper is organized as follows. Section \ref{sec_Prelim}, outlines the GSR problem and reviews Bayesian CNNs. Section \ref{sec_BCNNs} describes the architecture and statistical interpretation of the proposed CNNs. Section \ref{sec_learning} presents the BCNN-based prior learning algorithm. Section \ref{sec_VBGSR}, details the VB inference method for signal recovery. Discussion about the computational complexity and local convergence of the proposed algorithm are available in Sections \ref{sec:complexity} and \ref{sec:convergence}, respectively. Experimental results are reported in Section \ref{sec:Simulation} to validate the proposed algorithm. Finally, Section \ref{sec_concolusion} concludes the paper.

\section{Preliminaries}
\label{sec_Prelim}

In this section, we briefly explain the GSR problem and the Bayesian CNNs.

\subsection{Graph Signal Recovery}
\label{sec_GSR}
In this paper, we aim to recover a graph signal from under-sampled and noisy observations. Consider an undirected graph $G=(\cal V,\cal E)$ with a number of $N$ vertices, where $\cal V$$=\{v_1,v_2,...,v_N\}$ represents the set of vertices, and $\cal{E} \subset \cal{V}\times \cal{V}$ denotes the edges. We denote $\Wb$ as the adjacency matrix of the graph with the elements encoding the edge weights, and $\Lb=\Db-\Wb$ as the Laplacian matrix, where $\Db$ is the diagonal degree matrix with entries defined as $D_{m}=\sum_n W_{mn}$. The noisy under-sampled observation can be obtained by sampling the graph signal on the $M$ vertices that are selected at random, i.e., \cite{Shum13}, \cite{Ales24}, \cite{Chen15}
\begin{equation}
\label{eq: model}
\yb=\Psi\xb+\nb,
\end{equation}
where $\Psi\in\mathbf{R}^{M\times N}$ is the sampling matrix, containing a ``one'' entry at each row corresponding to the sampled node, and other entries equal ``zero'', $\xb\in\mathbf{R}^{N\times 1}$ denote the graph signal, and $\nb\in\mathbf{R}^{M\times 1}$ is the measurement noise.

In this paper, we consider the problem of recovering the original graph signal $\xb$ from a subset of nodes corrupted by noise. A statistical GSR algorithm based on variational Bayesian inference algorithm is used in this work to estimate the graph signal from observations.

\subsection{Deep Neural Networks and Gibbs Distribution}
\label{sec_DNNGibbs}
In this paper, we aim to solve the GSR problem using DNNs. Considering that the traditional DNNs can be thought of as a function that maps a high-dimensional input to a low-dimensional output, one can design a DNN for graph signal processing by retrieving standard DNNs. More precisely, DNNs first derive an initial retrieval $\xb^P$ using fully connected layers and, then, refine $\xb^P$ to derive a final retrieval $\hat{\xb}$ by exploiting the convolutional layers.

DNN outputs are deterministic given fixed weights and inputs. However, we often interpret the output logits (before softmax) or the softmax outputs (after transformation) as expressing probabilistic beliefs about the classes. We model DNNs as Gibbs distributions not because they are probabilistic by nature, but because doing so enables a rich probabilistic interpretation of their outputs, useful in uncertainty estimation, decision theory, and connections to physics and Bayesian inference. The authors of \cite{Lan19} have established a novel statistical framework based on the Bayesian inference for understanding the structure of DNNs. In particular, they have proven that the hidden layers of DNNs can be modeled by the Gibbs distributions of network input. A Gibbs distribution (also called Boltzmann distribution) is often used in mathematics and statistical mechanics to statistically describe the probability of a certain state as a function of system temperature and the energy of that state, and for the signal vector $\xb$ is given by \cite{Geman84}
\begin{equation}
p_i(\xb,\theta)=\frac{1}{Z(\theta)}\exp\left(-\frac{E_i(\xb,\theta)}{T}\right),
\end{equation}
where $\theta$ denotes all parameters, $Z(\theta)$ indicates the partition function, written as $Z(\theta)=\sum_{i=1}^I \exp\left(-\frac{E_i(\xb,\theta)}{T}\right)$, T indicates the temperature of system, $E_i(\xb,\theta)$ denotes the energy function for the state i, and $I$ is the total number of states in the system.

The Hammersley-Clifford theorem \cite{Chand17} states that the Gibbs distribution is equivalent to the joint probability distribution of any Markov random field (MRF). Any MRF can be written as exponential family
\begin{equation}
p(\xb,\theta)=\frac{1}{Z(\theta)}\exp\left(-\frac{E(\xb,\theta)}{T}\right),
\end{equation}
where, the energy function of the MRF, $E(\xb,\theta)$ defined as \cite{Zhu97}
\begin{equation}
E(\xb,\theta)=\sum_{i=1}^I [g_i^{NL}(f_i(\xb);\theta_i)],
\end{equation}
where $g_i^{NL}(.)$ is a non-linear function, $g_i^{NL}(f_i(.))$ may be referred to as clique potentials, and $f_i(.)$ is a linear filter.

In the literature, the connections between Gibbs distribution (particularly MRF) and DNNs is discussed \cite{Bour89}, \cite{Wu16}. Especially, the authors of \cite{Brid90}, \cite{LeCun06} proved that the softmax layer of DNNs is the same as Gibbs distribution. Moreover, it have proven that the MRF model can be used to formulate the convolutional layer containing a non-linear layer \cite{Zheng15}, \cite{Lan18}.

\section{The architecture of the Bayesian CNNs}
\label{sec_BCNNs}
In previous section, we have briefly discussed the connections between DNNs and Gibbs distribution. On the other hand, the main purpose of our work is recovering graph signal from noisy undersampled observations. But, the denoising ability of the popular activation functions of DNNs is weak \cite{Lan19}; which leads to poor performance of DNNs for GSR problem. To overcome this limitation, we propose a CNNs-based algorithm, which uses new activation function to improve the robustness of the DNNs against noise.

In this paper, we propose a new CNNs to learn the prior distribution $p(\xb;\theta)$; and, then, based on the proposed model for the CNNs, we derive the posterior distributions of hidden variables using the VB inference method. The proposed activation function improves robustness of DNNs \cite{Lan19}.

\begin{figure}[!t]
\centering
{\includegraphics[width=.95\linewidth]{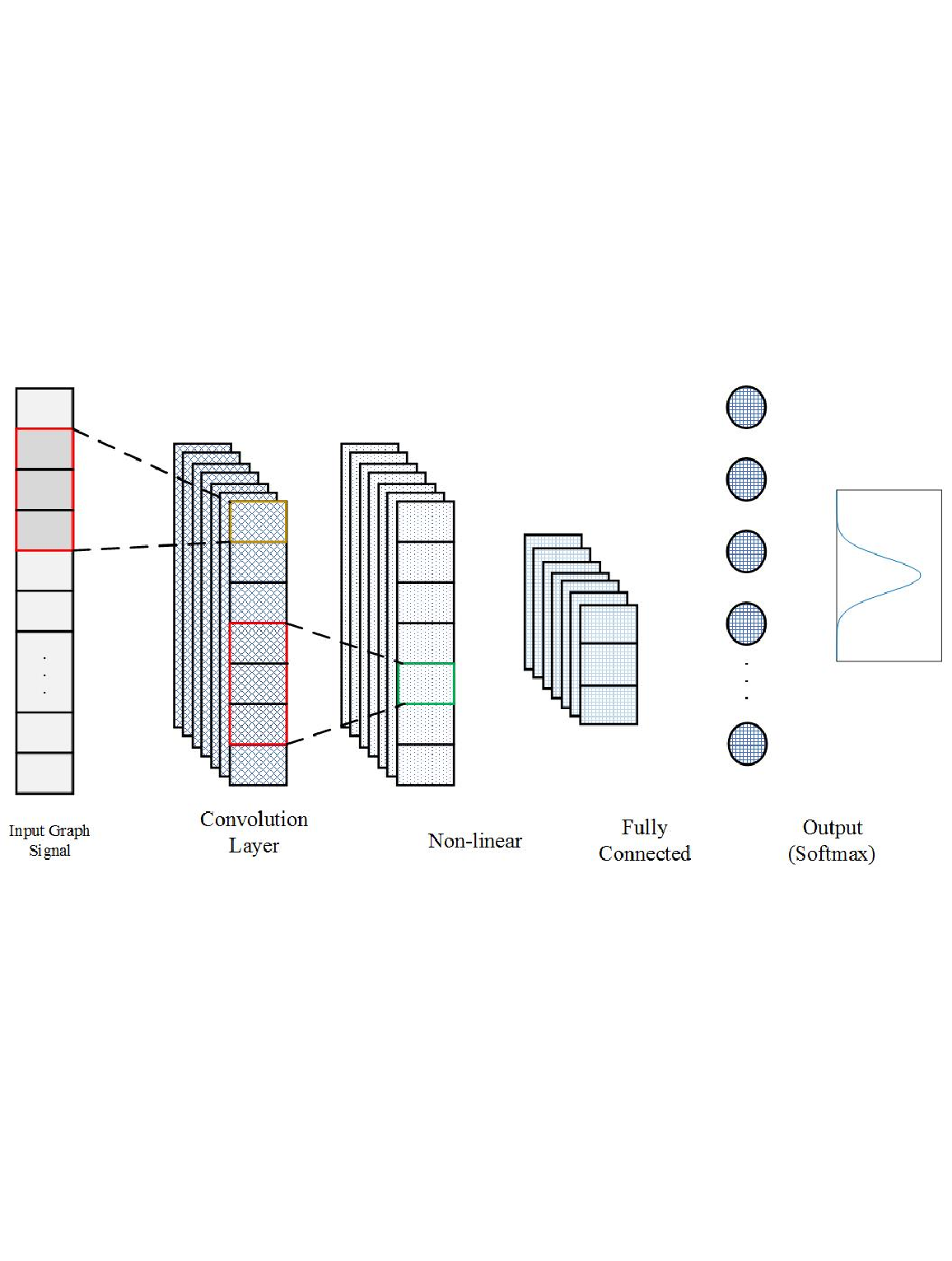}}%
\caption{Block diagram of the proposed CNNs for modeling a prior distribution $p(\xb;\theta)$ for the graph signal.}
\label{fig2}
\end{figure}

The structure of the proposed CNNs, shown in Fig.\ref{fig2}, consists of three hidden layers: convolutional, nonlinear (activation function), and fully connected layers. We assume the output layer to be softmax. Since the hidden layer of the proposed CNNs, is a fully connected layer, thus, we can assume that there are a number of $K$ states to write a Gibbs distribution formula for $\xb$. Note that the output of a convolutional filter is not a scalar to represent the weight of a state, but it is a matrix with high dimensions. Thus, the discrete Gibbs distribution is not able to provide a suitable representation for the convolutional layer. As an alternative, we can formulate the convolutional layer using MRF model as follows
\begin{equation}
p(\xb;\theta)=\frac{1}{Z(\theta)}\exp\{\sum_{m=1}^M [g_m^{NL}(f_m(x);\theta_m)]\},
\end{equation}
where ${Z(\theta)}$ is the partition function, $\sum_{m=1}^M [g_m^{NL}(f_m(x);\theta_m)]$ denote the fully connected layer, $g_m^{NL}$ is the non-linear layers (activation function) and $f_m$ denote the spectral graph convolutions.

Similar to \cite{Lan19}, we chose $g_m^{NL}$ to be as GMM model, which has better robustness against noise than the other activation functions \cite{Ma16}, \cite{Portallia03}. Thus, we can write
\begin{equation}
g_m^{NL}=\log \left[\sum_{n=1}^N \pi_{mn}\mathcal{N}(f_m(.);0,\sigma^2_{mn})\right],
\end{equation}
where $\pi_{mn}$ is the mixing weight and $\sigma^2_{mn}$ is the variance of each Gaussian in the GMM distribution, respectively. Moreover, we incorporate graph convolution operators into the CNN architecture so that the learned filters respect the topology of the underlying graph. To efficiently perform convolution on graph-structured data, we employ Chebyshev polynomial filters \cite{Deff16}, which provide a fast and localized approximation of spectral graph convolutions. These filters are based on Chebyshev polynomials of the rescaled graph Laplacian, which is efficient and well-suited for spectral graph filtering, enabling the design of $K$-localized filters that operate on a node's $K$-hop neighborhood. The convolution of a signal $\xb \in \mathbf{R}^{N}$ with the $m'$th Chebyshev polynomial filter of order $P$ is:
\begin{equation}
f_m(\xb)=\sum_{p=0}^P \beta_{m,p} T_p(\tilde{\Lb})\xb
\end{equation}
where $\tilde{\Lb}=\frac{2}{\lambda_{max}}\Lb-I$ is the scaled Laplacian, $\lambda_{max}$ is the largest eigenvalue of $\Lb$, $\beta_{m,p}$ are trainable coefficients (like filter weights), and $T_p$ are Chebyshev polynomials recursively defined as $T_0(x)=1$, $T_1(x)=x$, and $T_p(x)=2xT_{p-1}(x)-T_{p-2}(x)$ for $p\leq 2$. This approach avoids the need for explicit eigendecomposition of the Laplacian, significantly reducing computational complexity while maintaining the ability to model complex graph signal structures. The use of Chebyshev filters ensures that the learned convolutional operations respect the graph topology, making them well-suited for graph signal recovery tasks and graph neural networks.

Thus, the resulting prior distribution for the graph signal in the proposed CNNs is
\begin{equation}
p(\xb;\theta)=\frac{1}{Z(\theta)}\prod_{m=1}^M \sum_{n=1}^N \pi_{mn} \mathcal{N}(f_m(\xb);0,\sigma^2_{mn}),
\end{equation}

\begin{algorithm}[tb]
\caption{CNNs based prior learning algorithm}
\textbf{Input}   \textbf{Training signals}; \textbf{CNNs model}; $\mathrm{max_{-}iter}$. \newline
\textbf{Initialize} GMM variances $\sigma_{mn}$, CNNs parameters $\{f,\pi\}$, sampling iteration, training iteration $t=0$, learning rate, sub-signal size.
\label{Algorithm_1}
\begin{algorithmic}
\REPEAT
\begin{itemize}
\item Generate training sub-signals from training signals
\item Compute $\left<\frac{\partial \log p(x;\theta)}{\partial\theta}\right>_{p(x_k^{data};\theta_t )}$ using training sub-signals
\item Get samples from $\left<\frac{\partial \log p(x;\theta)}{\partial\theta}\right>_{p(x_k^{prior};\theta_t)}$ to obtain training labels
\item Compute $\left<\frac{\partial \log p(x;\theta)}{\partial\theta}\right>_{p(x_k^{prior};\theta_t)}$ using training labels
\item Update the parameter set $\theta_t$ using CD algorithm.
\item $t \gets t+1$
\end{itemize}
\UNTIL $|\theta_{t+1}-\theta_t|<\epsilon$ or $t>\mathrm{max_{-}iter}$
\end{algorithmic}
\end{algorithm}
\vspace{-2ex}

In summary, in the proposed CNNs, the prior distribution is formulated, where the features are learned by convolution layer from the training dataset, and in order to introduce nonlinearity and robustness against noise, the GMM model is exploited for the activation function.

\section{The proposed prior learning algorithm based on BCNNs}
\label{sec_learning}
In previous section, we have defined the CNNs architecture. Now we aim to design an algorithm to learn the CNNs parameters by optimizing $\theta=\{f,\pi\}$, in which $f=[f_1,...,f_M]$ and $\pi=[\pi_{11},...,\pi_{NM}]$. To this end, we measure the distance between $p(x^{prior};\theta)$ and $p(x^{data};\theta)$ using Kullback-Leibler divergence (KLD) \cite{Hint02} criteria. The KLD is defined as
\begin{align}
KLD(p&(x^{prior};\theta)||p(x^{data};\theta)) \nonumber \\
&=\sum_i p(x^{data}(i);\theta).\log \frac{p(x^{data}(i);\theta)}{p(x^{prior}(i);\theta)},
\end{align}
where $p(x^{prior};\theta)$ is the distribution resulted from sampling $p(x;\theta)$, and $p(x^{data};\theta)$ is the output of the proposed CNNs, i.e., the distribution of the training data. We expect $p(x;\theta)$ to model the statistical behaviour of the training dataset accurately. To this end, we exploit the gradient descent algorithm \cite{Hint02} to minimize the KLD.

Similar to \cite{Lan19}, we use the contrastive divergence (CD) learning method \cite{Carr05} for optimizing $\theta=\{f,\pi\}$, which only takes a few sampling iterations to estimate $p(x;\theta)$, and, thus, increase the training speed of the optimizing algorithm. The CD algorithm can be formulated as
\begin{align}
&\theta_{n+1}=\theta_n\nonumber\\
&-\eta\left[\left<\frac{\partial \log p(x;\theta)}{\partial\theta}\right>_{p(x_k^{prior};\theta_n)} -\left<\frac{\partial \log p(x;\theta)}{\partial\theta}\right>_{p(x_k^{data};\theta_n )}\right],
\end{align}
where $\theta_n$ represents the optimized parameters in the $n-$th training, $\left<\right>$ denotes statistical averaging with respect to its indices, and $\eta \in (0,1]$ represents the learning rate. The training algorithm is summarized in Algorithm\ref{Algorithm_1}.

\section{The proposed VB inference algorithm for GSR}
\label{sec_VBGSR}
Given the noisy graph observations, i.e., $\yb=\Psi\xb+\nb$, Bayesian GSR converts graph signal recovery problem to a Bayesian posterior inference problem. It is a common assumption that the noise vector $\nb$ selected from a zero-mean white Gaussian noise with unknown precision (inverse of variance) $\alpha_e$. So, we have $p(\nb)={\cal N}(0,\alpha^{-1}_e\Ib_{M})$. Thus, we can write the Likelihood function as
\begin{equation}
\label{eq_likelihood}
p(\yb|\Psi,\xb,\alpha^{-1}_e)={\cal N}(\Psi\xb,\alpha^{-1}_e\Ib_{M}).
\end{equation}

We unknown variables are denoted by $\thetab=\{\xb, \alpha_e\}$. If we denote $p(\theta_i;\theta)$ to be the prior distribution of the unknown variable $\theta_i$, then, using the Bayes' theorem, the posterior distribution can be derived as
\begin{equation}
p(\theta_i|\Psi,\yb,\thetab\setminus\theta_i)\propto p(\yb|\Psi,\xb,\alpha^{-1}_e).p(\theta_i;\theta),
\end{equation}
where $\thetab\setminus\theta_i$ means all unknown variables except $\theta_i$

In this section, we use variational Bayesian framework to infer the posterior distributions based on the prior distributions of the unknowns and the likelihood distribution $p(\yb|\Psi,\xb,\alpha^{-1}_e)$. The prior distribution for the graph signal, $p(\xb;\theta)$, is proposed in section \ref{sec_BCNNs}. Moreover, we assume a Gamma distribution as the prior for the noise precision $\alpha_e$, i.e., $p(\alpha_e)=\mathrm{Gamma}(\rho_e^0,\xi_e^0)$, which is the conjugate of Gaussian distribution, and leads to a closed form statement of the posterior distribution.

In VB inference method, the posterior $p(\thetab|\yb)$ is not calculated; but, instead, an estimate of it, which we denote by $q(\thetab)$, is inferred in an iterative procedure, such that the KLD between the posterior $p(\thetab|\yb)$ and the estimated posterior $q(\thetab)$ is minimized \cite{Beal03}. In general, we have \cite{Tork21},\cite{Tork24}
\begin{equation}
\ln(q(\theta_i))= \left<\ln(p(\thetab),\yb)\right
>_{\thetab\setminus\theta_i}+\mathrm{const.},
\end{equation}
where $\left<\right>$ denotes statistical averaging with respect to the indices. We also assume factorization in the posterior distributions; i.e.,
\begin{equation}
q(\theta)= q(\xb)q(\alpha_e).
\end{equation}

Based on the prior distribution described in section \ref{sec_BCNNs} and the Likelihood function presented in \eqref{eq_likelihood}, we calculate the posterior distribution for the graph signal $\xb$ as
\begin{align}
\ln q(\xb)=& \left<\ln p(\yb, \xb, \alpha_e)\right>_{\alpha_e}+\mathrm{const.} \nonumber \\
=&\left<\sum_{m=1}^M \ln\left[\sum_{n=1}^N \pi_{mn} \mathcal{N}(f_m(\xb);0,\sigma^2_{mn})\right]\right>_{\alpha_e}\nonumber \\
&-\left<\frac{\alpha_e}{2}(\yb-\Psi\xb)^{T}(\yb-\Psi\xb)\right>_{\alpha_{mn}, \alpha_e}+\mathrm{const.}\nonumber \\
=&\left<\sum_{m=1}^M \ln\left[\sum_{n=1}^N \pi_{mn}\exp\left(\frac{-\xb^Tf_m^Tf_m\xb}{2\sigma^2_{mn}}\right)\right]\right>_{\alpha_e}\nonumber\\
&-\left<\frac{\alpha_e}{2}(\xb^T\Psi^T\Psi\xb-2\xb^T\Psi^T\yb)\right>_{\alpha_e}+\mathrm{const.}\nonumber\\
=&<\sum_{m=1}^M \{\ln\left[\sum_{n=1}^N \pi_{mn}\exp\left(\frac{-\xb^Tf_m^Tf_m\xb}{2\sigma^2_{mn}}\right)\right]\nonumber\\
&-\frac{\alpha_e}{2M}(\xb^T\Psi^T\Psi\xb-2\xb^T\Psi^T\yb)\}>_{\alpha_e}+\mathrm{const.},
\end{align}

Thus, the posterior for the graph signal is
\begin{align}
q(\xb)\propto&\prod_{m=1}^M \ln[\sum_{n=1}^N \pi_{mn}\exp(\frac{-\xb^Tf_m^Tf_m\xb}{2\sigma^2_{mn}}\nonumber\\
&+\frac{-\xb^T\Psi^T\Psi\xb+2\xb^T\Psi^T\yb}{2M\sigma^2_e})]\nonumber\\
=&\frac{1}{Z'(\theta)}\prod_{m=1}^M \sum_{n=1}^N \pi'_{mn} \mathcal{N}(\mu_{mn},\Sigma_{mn}),
\end{align}
which belongs to the GMM distribution with $Z'(\theta)$ denoting the partition function and the following parameters
\begin{align}
\label{eq_signal}
&\Sigma_{mn} =\left(\frac{f_m^T f_m}{\sigma^2_{mn}}+\frac{\Psi^T \Psi}{M\sigma^2_e}\right)^{-1},\nonumber\\
&\mu_{mn} =\Sigma_{mn} \frac{\Psi^T\yb}{M\sigma^2_e},\nonumber\\
&\pi_{mn}'=\frac{1}{Z_{\pi}}\pi_{mn}\exp\left(\frac{\mu_{mn}^T\Sigma_{mn}^{-1}\mu_{mn}}{2}-\frac{\yb^T\yb}{2\sigma^2_e}\right),
\end{align}
where the parameters $\sigma^2_{mn}$ and $f_m$ are derived in the prior learning process (Sec. \ref{sec_learning}), and $Z_{\pi}$ is the partition function for $\pi_{mn}'$, defined as
\begin{displaymath}
Z_{\pi}=\sum_{m=1}^M {\sum_{n=1}^N {\pi_{mn}\exp\left(\frac{\mu_{mn}^T\Sigma_{mn}^{-1}\mu_{mn}}{2}-\frac{\yb^T\yb}{2\sigma^2_e}\right)}}
\end{displaymath}

The posterior distribution for the precision of the noise can be derived as
\begin{align}
\ln q(\alpha_e)=& \left<\ln p(\yb, \xb, \alpha_e)\right>_{\xb}\nonumber\\
=&\left<-\frac{\alpha_e}{2}(\yb-\Psi\xb)^{T}(\yb-\Psi\xb)\right>_{\xb}\nonumber\\
&-\xi_e^0\alpha_e+(\rho_e^0-1)\ln(\alpha_e) \nonumber \\
&-\xi_e^0 \alpha_e+\frac{M}{2}\ln(\alpha_e)+\mathrm{const.},
\end{align}

\begin{algorithm}[tb]
\caption{Proposed BCNNs algorithm for graph signal recovery}
\textbf{Input}   \textbf{Observations} $\yb$; \textbf{Sampling matrix} $\Psi$; $p(\xb|\theta)$; $\mathrm{max_{-}iter}$; $\epsilon$. \newline
\textbf{Initialize} $\hat{\xb}=\Psi^{T}\yb, \rho_e^0, \xi_e^0$, $t=0$.
\label{Algorithm_2}
\begin{algorithmic}
\REPEAT
\begin{itemize}
\item Update the noise precision parameters $\rho_e$ and $\xi_e$ using \eqref{eq_noiseparam}
\item Update the mean of the noise precision $q(\alpha_e)$ by $\left<\alpha_e\right>=\frac{\rho_e}{\xi_e}$ and then set $\sigma^2_e=\alpha^{-1}_e$
\item Use \eqref{eq_signal} to update the parameters of posterior of graph signal
\item Set $\hat{\xb}=\mu$ and compute $\mathrm{error}=\Vert \xb-\hat{\xb}\Vert_{2}^2/\Vert \xb\Vert_{2}^2$ and $t\leftarrow t+1$
\end{itemize}
\UNTIL $\mathrm{error}<\epsilon$ or $n\leq \mathrm{max_{-}iter}$
\end{algorithmic}
\end{algorithm}
\vspace{-2ex}

It is evident that $\alpha_e$ has Gamma distribution given by
\begin{align}
\label{eq_noiseparam}
q(\alpha_e)&=\mathrm{Gamma}(\rho_e,\xi_e), \nonumber \\
\rho_e&=\rho_e^0+\frac{M}{2}, \nonumber \\
\xi_e&=\xi_e^0+\frac{1}{2}\left<(\yb-\Psi\xb)^T(\yb-\Psi\xb)\right>_{\xb} \nonumber \\
&=\xi_e^0+\frac{1}{2}(\yb-\Psi\mu)^T(\yb-\Psi\mu)+\frac{1}{2}\mathrm{Tr}(\Psi\Sigma\Psi^T).
\end{align}

The graph signal recovery algorithm is summarized in Algorithm \ref{Algorithm_2}.
\section{Computational complexity}
\label{sec:complexity}
This section presents the computational complexity of the proposed BCNN-GSR algorithm. Note that the algorithm requires training a CNN prior, with M convolutional filters, GMM-based nonlinearities, and optimization using contrastive divergence (CD). Then, the algorithm uses VB inference with learned prior. Inference involves multiple mixture components, learning/using GMM weights $\pi_{mn}$ and variances $\sigma_{mn}$, and update expressions. The complexity of training the CNN is $\mathcal{O}(NKMT)$ where $T$ is the number of iterations. In the inference procedure, the GMM-based posterior updates involve summing over multiple components, and the computational complexity is about $\mathcal{O}(M.N^2)$ for each VB step.

\section{Local convergence of the proposed BCNN-GSR algorithm}
\label{sec:convergence}
In this section, we investigate the convergence of the proposed VB-based GSR algorithm. To this end, we rewrite the update equations for hidden variables derived in Sec. \ref{sec_VBGSR}. Similar to the expectation-maximization (EM) algorithm, we first optimize hyperparameters at iteration $t$, where we have the posteriors of $(t-1)$'th iteration. This gives
\begin{align}
\label{eq_noiseparamk}
\rho_e^{(t)}=&\rho_e^0+\frac{M}{2}, \nonumber \\
\xi_e^{(t)}=&\xi_e^0+\frac{1}{2}(\yb-\Psi\mu^{(t-1)})^T(\yb-\Psi\mu^{(t-1)})\nonumber \\
&+\frac{1}{2}\mathrm{Tr}(\Psi\Sigma^{(t-1)}\Psi^T).
\end{align}

Then, we optimize the posteriors based on the hyperparameters calculated in \eqref{eq_noiseparamk} as follows
\begin{align}
\label{eq_posteriork}
&q^{(t)}(\alpha_e)=\mathrm{Gamma}(\rho_e^{(t)},\xi_e^{(t)}), \nonumber \\
&q^{(t)}(\xb)=\frac{1}{Z^{(t)}(\theta)}\prod_{m=1}^M \sum_{n=1}^N \pi^{(t)}_{mn} \mathcal{N}(f_m(\xb);\mu^{(t)}_{mn},\Sigma^{(t)}_{mn}),
\end{align}
where
\begin{align}
&\Sigma^{(t)}_{mn} =\left(\frac{f_m^T f_m}{\sigma^2_{mn}}+\frac{\Psi^T \Psi}{M(\sigma^2_e)^{(t)}}\right)^{-1},\nonumber\\
&\mu^{(t)}_{mn} =\Sigma^{(t)}_{mn} \frac{\Psi^T\yb}{M(\sigma^2_e)^{(t)}},\nonumber\\
&\pi_{mn}^{(t)}=\frac{1}{Z^{(t)}_{\pi}}\pi_{mn}\exp\left(\frac{(\mu^{(t)}_{mn})^T(\Sigma^{(t)}_{mn})^{-1}\mu^{(t)}_{mn}}{2}-\frac{\yb^T\yb}{2(\sigma^2_e)^{(t)}}\right),
\end{align}

Note that the VB estimate of a parameter is the mean of estimated posterior distribution; i.e., $E[q(.;\theta)]$. Thus, the VB approximations at iteration $t$ are given by
\begin{align}
\label{eq_meank}
&\alpha_e^{(t)}=\frac{\rho_e^{(t)}}{\xi_e^{(t)}}=A(\theta^{(t-1)}), \nonumber \\
&\xb^{(t)}=\frac{1}{Z^{(t)}(\theta)}\prod_{m=1}^M \sum_{n=1}^N \pi^{(t)}_{mn} \mu^{(t)}_{mn}=B(\theta^{(t-1)}),
\end{align}

Thus, the above procedure can be rewritten as
\begin{equation}
\label{eq_conver}
\theta^{(t)}=\Phi(\theta^{(t-1)})=
\begin{bmatrix}
A(\theta^{(t-1)}) \\
B(\theta^{(t-1)})
\end{bmatrix},
\end{equation}

Similar to the proof presented in \cite{Wang03}, we can show that the $\Phi(\theta)$ is locally contractive, i.e., there exist a number $\lambda, 0\leq\lambda\leq 1$, such that
\begin{equation}
||\Phi(\theta^*)-\Phi(\bar{\theta})||\leq \lambda ||\theta^*-\bar{\theta}||
\end{equation}
where $\theta^*$ is the true value of $\theta$, and $\bar{\theta}$ situates sufficiently near $\theta^*$. Thus, we can write
\begin{align}
||\theta^{(t)}-\theta^*||&\leq ||\Phi(\theta^{(t-1)})-\Phi(\theta^*)||+||\Phi(\theta^*)-\theta^*||\nonumber \\
&\leq \lambda ||\theta^{(t-1)}-\theta^*||+||\Phi(\theta^*)-\theta^*||,
\end{align}
which demonstrates that as the iteration number $t$ increases, $\Phi(\theta^*)$ in the iterative procedure \eqref{eq_conver} tends to the true value $\theta^*$ locally.

\section{Simulation results}
\label{sec:Simulation}
In this section, we present simulation results that confirm the performance of the proposed BCNN-GSR algorithm. First, we determine the effect of different choices of hyper-parameters on the proposed BCNNs model. Two hyper-parameters appear in the proposed BCNNs model: $F_n$ which denotes the number of filters $f_m$, i.e., , and GMM variances. After choosing the better set of parameters, we evaluate the performance of prior learning algorithm for different types of probability distributions of graph signal. Finally, we demonstrate the graph signal recovery performance of our proposed algorithm. We use normalized mean square error (NMSE) criteria to compare different choices of hyper-parameter sets and different algorithms, which is calculated as
\begin{displaymath}
\mathrm{NMSE}=\frac{1}{K}\sum_{k=1}^K \frac{||\hat{\xb}_k-\xb_k||^{2}_2}{||\xb_k||^{2}_2}.
\end{displaymath}

\begin{figure}[!t]
\centering
{\includegraphics[width=.95\linewidth]{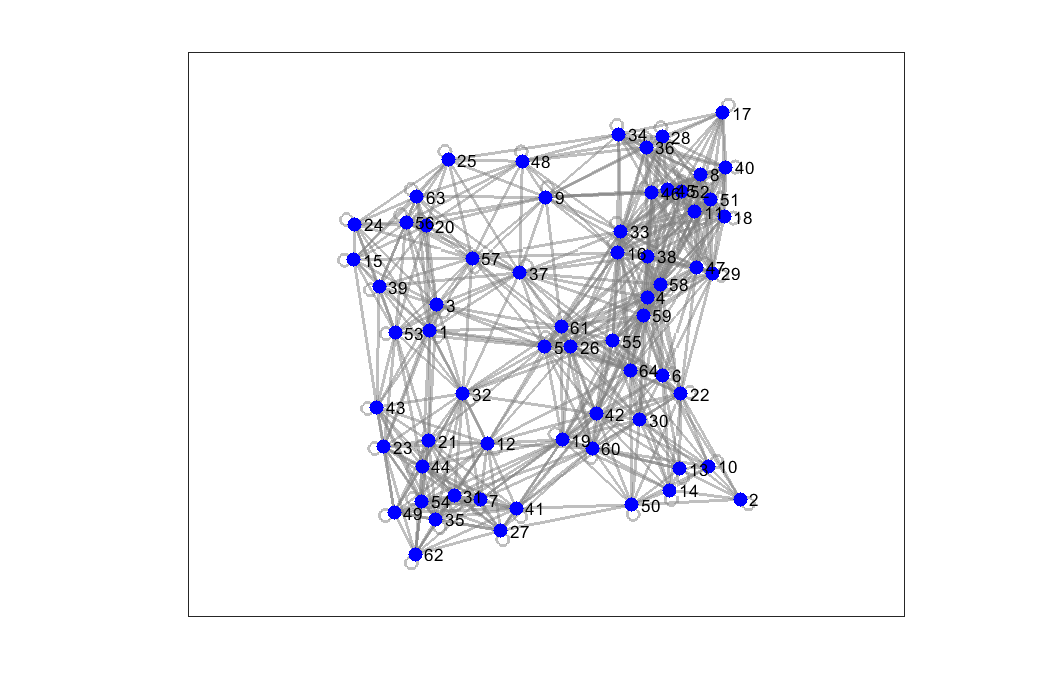}}%
\caption{The original graph with $N=64$ nodes.}
\label{fig3}
\end{figure}

The proposed BCNN-GSR algorithm is tested using a synthetic graph with $N=64$ nodes, where the coordinates of the vertices are randomly selected. To calculate the edge weights, Gaussian radial basis function (RBF) is used which is defined as $\mathrm{exp}\left(-\frac{d^2(i,j)}{2 \sigma^2}\right)$, with $d(i,j)$ denoting the distance between vertices $i$ and $j$, and $\sigma$ is the kernel width parameter. In the simulation, we select $\sigma=0.5$ and remove the edges with weight less than $0.75$. Finally, we compute the graph Laplacian matrix by $\Lb =\mathrm{diag}(\Wb .\onevec_N)-\Wb$ and normalize it with its trace. The graph is shown in Fig.\ref{fig3}.

\subsection{Choosing hyper-parameters}
\label{simul_hyper}
In this subsection, we determine the effect of different hyper-parameters on modelling the prior distribution of graph signal. To generate training graph signals, we consider two scenario: $(i)$ we use the band-limited graph signal model $\xb[k]=\sum_{i=1}^\omega \gamma_k^{(i)}\rb^{(i)},(k=1, \dots , K)$, where $\gamma_k^{(i)}\sim \mathcal{N}(0,1)$, $\omega$ is the bandwidth parameter, and $\{\rb^{(i)}\}_{i=1}^\omega$ denotes the eigenvector corresponding to the $\omega$ smallest eigenvalues of the Laplacian matrix; $(ii)$ we generate training graph signals from GMM distribution with the mixture of four Gaussian. In two scenarios, we generate $K=50$ graph signals with size $20\times 1$ as training dataset. Then, we extract $5 \times 1$ sub-signals from these graph signals for training. For the hyper-parameters set, we consider the convolutional depth $F_n$ to be equal 6 or 8. Moreover, we assume $P=3$ for the order of the Chebyshev polynomial. Similar to \cite{Lan19}, we choose the GMM variances from two sets $\sigma^{(1)}=\frac{0.001}{\delta^{(1)}}$ and $\sigma^{(2)}=\frac{0.001}{\delta^{(2)}}$; where $\delta^{(1)}=\exp\{-7,-3,0,3,7\}$ and $\delta^{(2)}=\exp\{\pm 7, \pm 5, \pm 3, \pm 1\}$. This set of hyper-parameters are summarized in Table \ref{Table_1}.

\begin{table}[!t]
\scriptsize
\centering
\caption{BCNNs models with different choices of hyper-parameters}
\begin{tabular}{l||p{1cm}|p{1cm}|p{1cm}} \hline
& & &   \\[-1.5ex]
 Model & $F_n$ & $\sigma$ \\ \hline
  BCNN1 & 6 & $\sigma^{(1)}$\\
  BCNN2 & 8 & $\sigma^{(1)}$\\
  BCNN3 & 8 & $\sigma^{(2)}$ \\\hline
\end{tabular}
\label{Table_1}
\end{table}

To evaluate the ability of different BCNNs models in estimating the graph signal probability distribution, we have compared the KLD between the empirical and the estimated distributions. The results for bandlimited GMRF with the bandwidth $\omega=25$ and the GMM distributions are shown in Table \ref{Table_2}. It is evident that the BCNN1 system with the convolutional depth $F_n=6$ has the highest KLD in both GMRF and GMM cases; i.e., its ability to estimate the prior distribution is the worst. Moreover, increasing $F_n$ from 6 (in BCNN1) to 8 (in BCNN2) model, gets smaller KLD. Finally, the system with $F_n=8$ and variance $\sigma^{(2)}$ in BCNN3 model, reaches smaller KLD. Briefly, the system with more hyper-parameters is able to better model the prior distribution. Thus, in subsequent simulations we use this system for modeling the prior distribution in graph signal recovery process.

\begin{table}[!t]
\scriptsize
\centering
\caption{The effect of different BCNNs hyper-parameters on modelling the prior distribution. }
\begin{tabular}{l||p{2cm}|p{2cm}} \hline
& &  \\[-1.5ex]
 Model & KLD (GMRF) & KLD (GMM) \\ \hline
  BCNN1 &      0.212 &0.207 \\
  BCNN2 &0.120 &0.113 \\
  BCNN3 & 0.031&0.024 \\\hline
\end{tabular}
\label{Table_2}
\end{table}

\begin{figure*}[!t]
\centering
\subfloat[NMSE for $\mathrm{SNR}=10dB$]{\includegraphics*[width=2.2in]{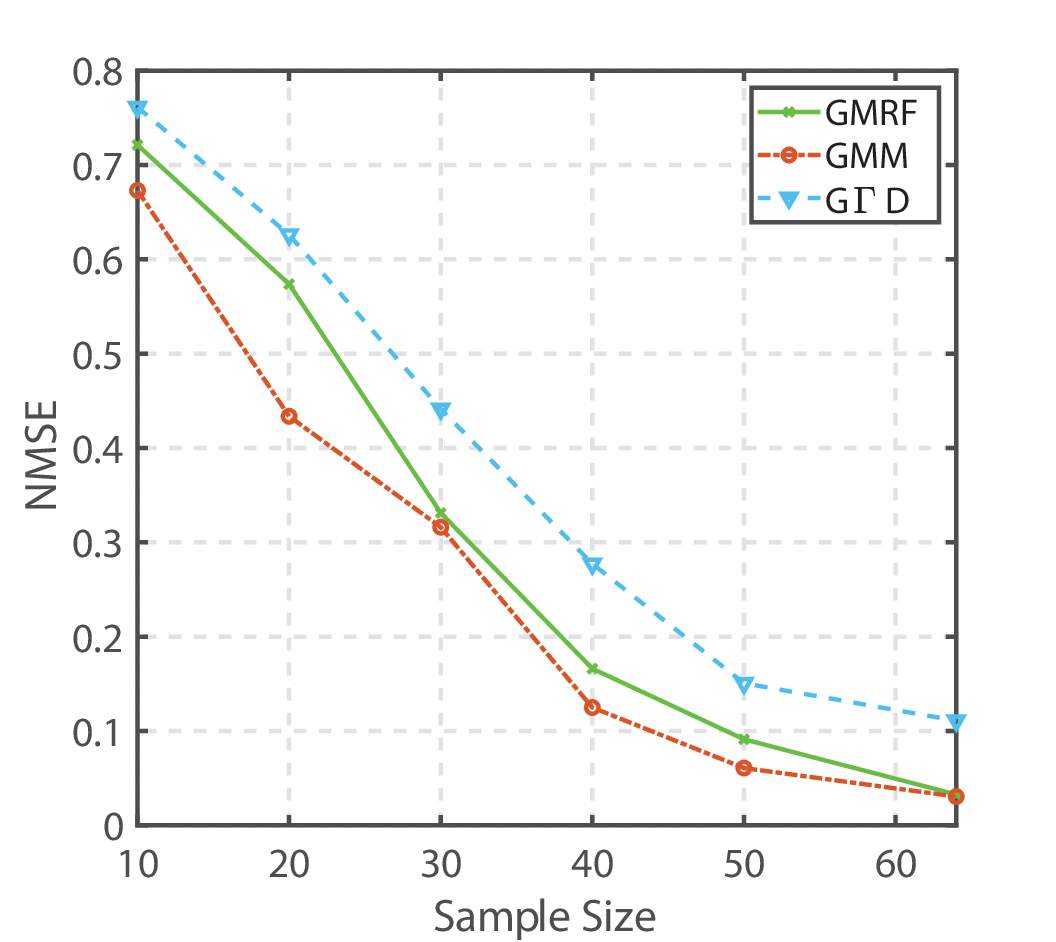}\label{BCNN10} } \hspace*{-0.3em}
\subfloat[NMSE for $\mathrm{SNR}=20dB$]{\includegraphics*[width=2.2in]{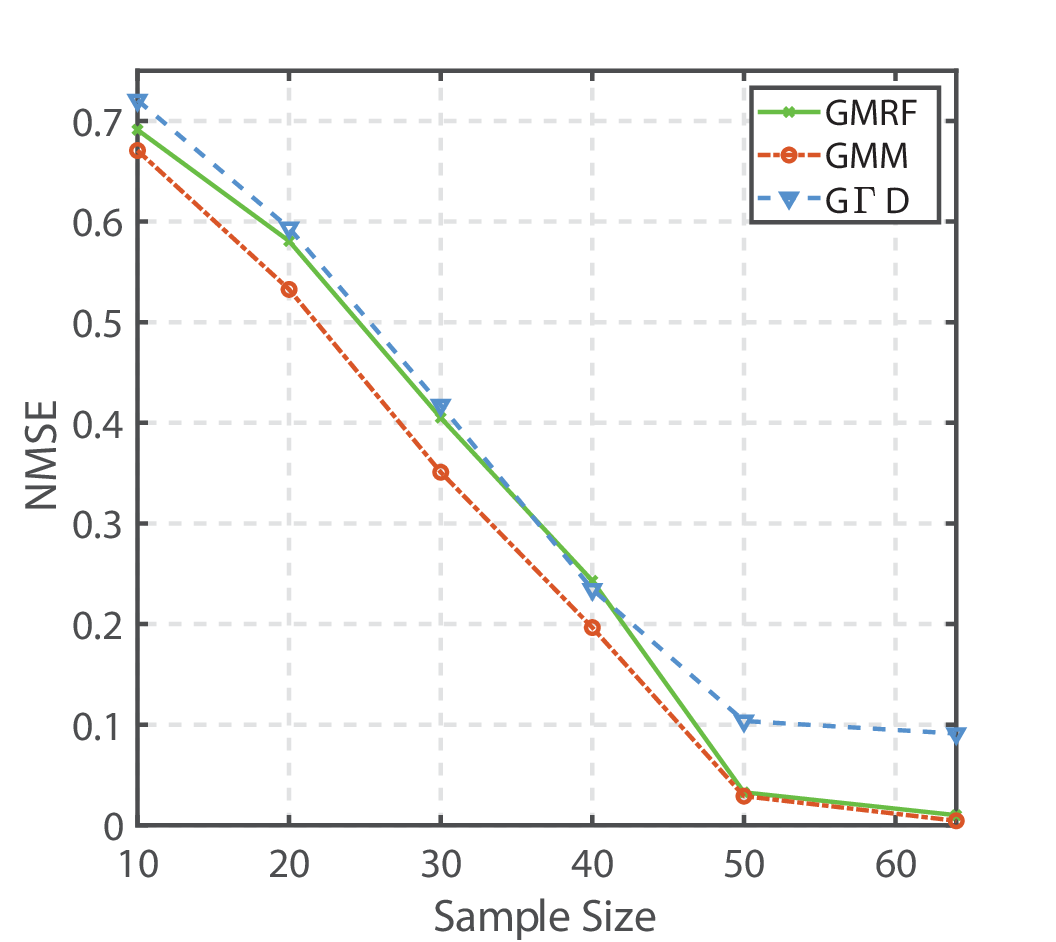}\label{BCNN20} }\hspace*{-0.3em}
\caption{The NMSE results of the BCNN-GSR for three different distributions.}
\label{BCNN_SNR}
\end{figure*}

\subsection{Performance of prior learning algorithm}
\label{simul_learn}
To confirm the ability of our proposed BCNNs-based learning algorithm in modeling the prior distribution and recovery process, the graph signals for both learning and recovery processes are generated from three different distributions: bandlimited GMRF, GMM, and generalized Gamma (G$\mathrm{\Gamma}$D) distributions. For generating graph signal with GMRF model, we used the same procedure as described in section \ref{simul_hyper}. For the GMM model, we generated a mixture of 4 Gaussian distributions. In all three scenarios, we generate $K=50$ graph signals with size $20\times 1$ as training dataset. Then, we extract $5 \times 1$ sub-signals from these graph signals for training. For the hyper-parameters, we have used BCNN3 model settings as presented in Table \ref{Table_1}. After estimating the prior distribution using the proposed CNNs algorithm, we exploit the learned prior in the VB inference procedure to recover the graph signal. As observation noise in recovery process, we add i.i.d. Gaussian noise to each graph signal, which causes $\mathrm{SNR}=10,20 dB$.

\begin{figure*}[!t]
\centering
\subfloat[NMSE for GMRF distribution]{\includegraphics*[width=2.2in]{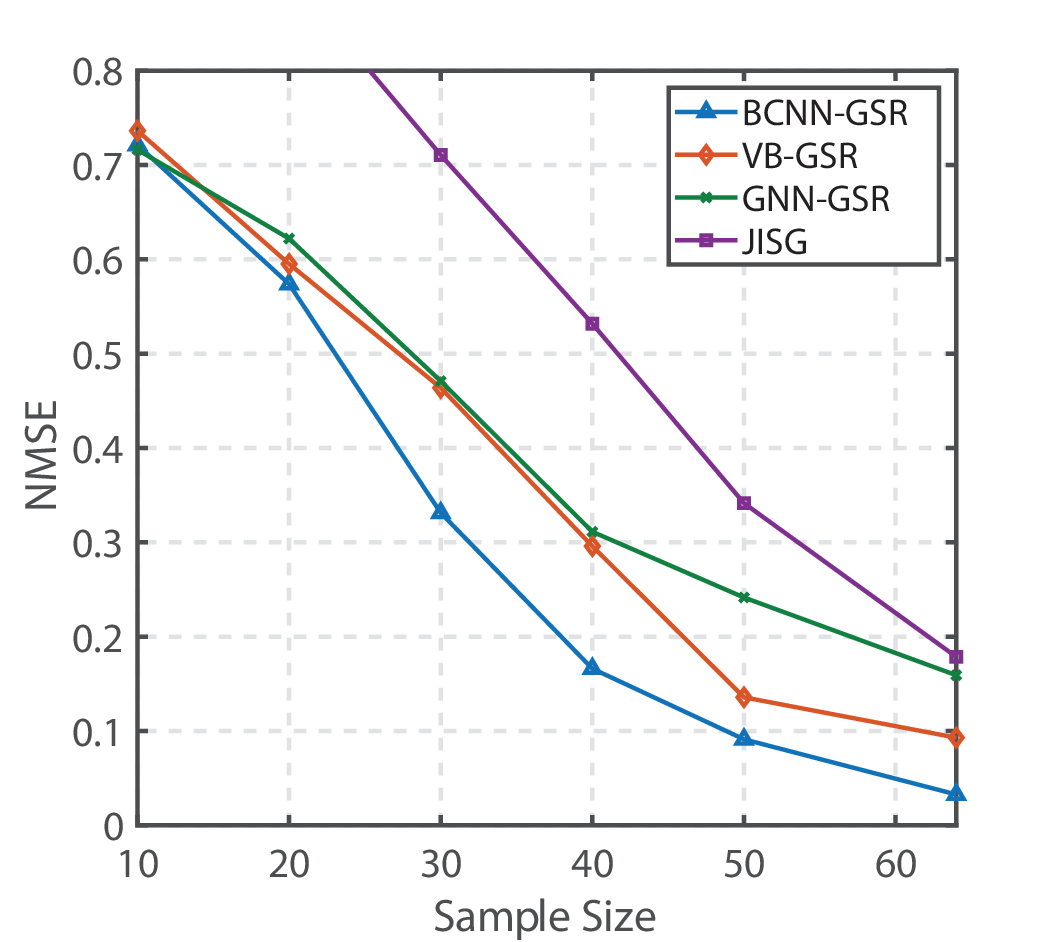}\label{GMRF10} } \hspace*{-0.3em}
\subfloat[NMSE for GMM distribution]{\includegraphics*[width=2.2in]{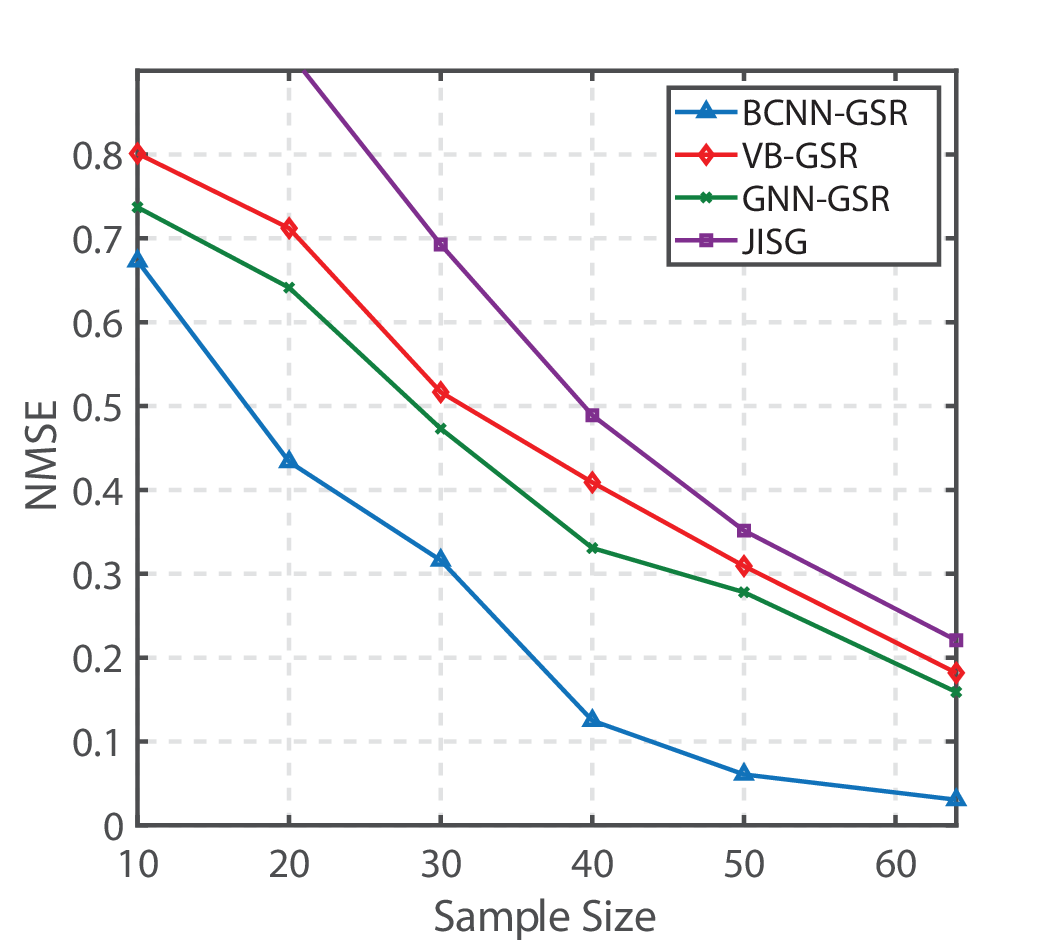}\label{GMM10} }\hspace*{-0.3em}
\subfloat[NMSE for G$\Gamma$D distribution]{\includegraphics*[width=2.2in]{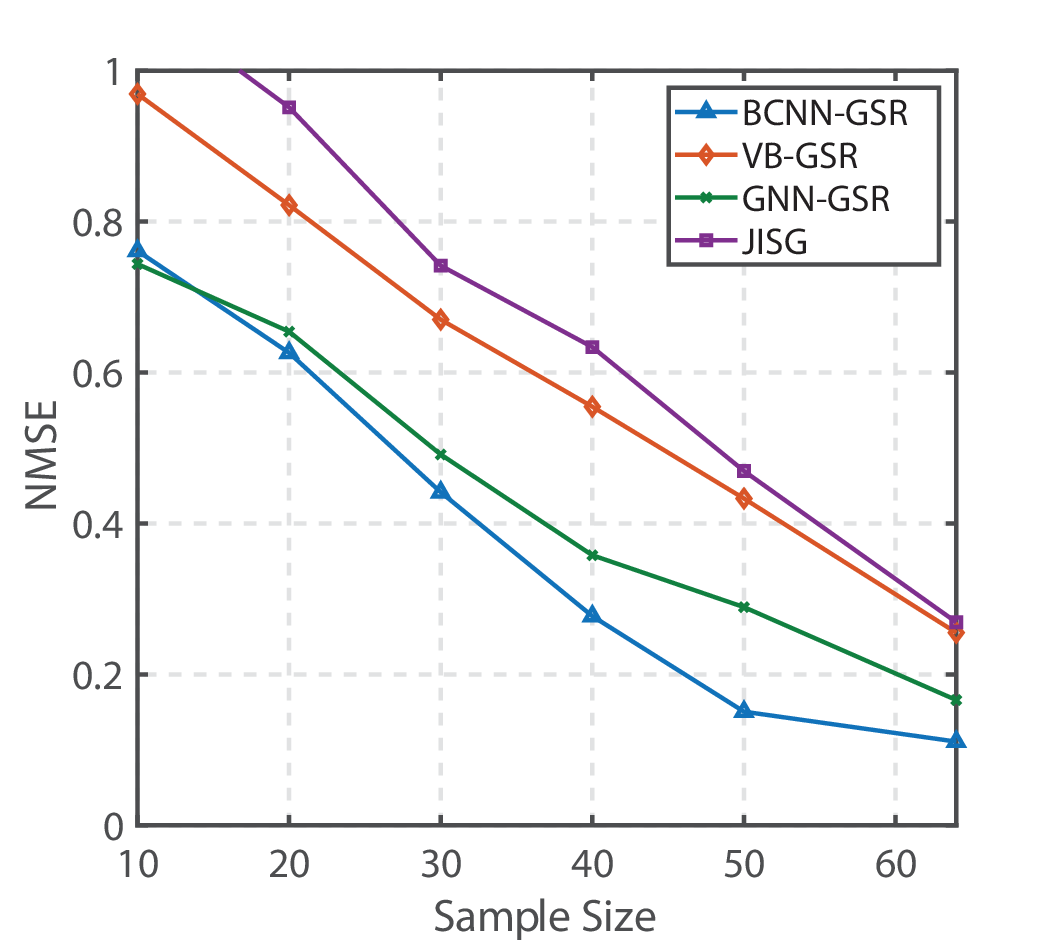}\label{GAMMA10} }\hspace*{-0.3em}
\caption{The NMSE comparison of different algorithms when the graph signal is generated from: (a) GMRF distribution; (b) GMM distribution; and (c) G$\Gamma$D distribution.}
\label{NMSE}
\end{figure*}

From the NMSE curves reported in Fig. \ref{BCNN_SNR}, it can be concluded that the proposed prior learning algorithm suitably models the statistical behaviors of the unknown original distribution. Moreover, when the original signal is generated from GMM family, i.e., GMRF and GMM models, the resulted NMSE is smaller than the other case; which proves the fact that the proposed GMM model for the non-linear layer of the CNNs, estimates the signals originated from GMM family better than the other distributions. Moreover, increasing SNR results in decreasing the NMSE for all distributions.

\subsection{Comparison with other GSR algorithms for Synthetic data}
\label{simul_GSR}
In this subsection, we compare the efficiency of our proposed GSR algorithm with some other methods for the synthetic graph signal generated as described in section \ref{simul_learn}. To evaluate the graph signal recovery performance of the proposed algorithm, we compare our method with three GSR algorithms: VB-GSR algorithm \cite{Tork21}, a statistical graph signal recovery algorithm based on variational Bayesian inference method, which assumes Gaussian MRF distribution as prior for all given graph signals; the GNN-GSR algorithm \cite{Chen20}, which uses graph neural networks for sampling and recovering graph signals; and non-dynamic part of Joint Inference of the network topology and Signals over Graphs (JISG) \cite{Ioan19}.

Fig. \ref{NMSE} shows the NMSE of different algorithms for different choices of prior distribution for graph signal. Fig. \ref{GMRF10} demonstrates that when the graph signal originated from the GMRF distribution, the performance of the proposed BCNN-GSR and VB-GSR algorithms is the same, which confirms the prominent ability of our proposed algorithm in learning the prior distribution. In other words, the VB-GSR algorithm assumes GMRF prior distribution for all graph signals, which conforms the original distribution; and the proposed BCNN-GSR algorithm estimates the prior distribution very well, such that the resulting NMSE is almost the same for the case where we already know the prior distribution and for the case where we estimate it.

From Figs. \ref{GMM10} and \ref{GAMMA10}, we conclude that when the graph signal originates from a probability distribution different from the assumption of the VB-GSR algorithm for prior distribution, the resulting NMSE of the recovered graph signal using VB-GSR algorithm is high. While, the proposed BCNN-GSR algorithm still works very well. This is due to the powerful ability of the CNNs-based prior learning algorithm, which estimates the statistical behaviour of the original graph signal very well, and exploits the resulting prior distribution in VB-inference procedure to recover the graph signal.

\begin{figure*}[!t]
\centering
\subfloat[Mean and variance for GMM signal]{\includegraphics*[width=5.2in]{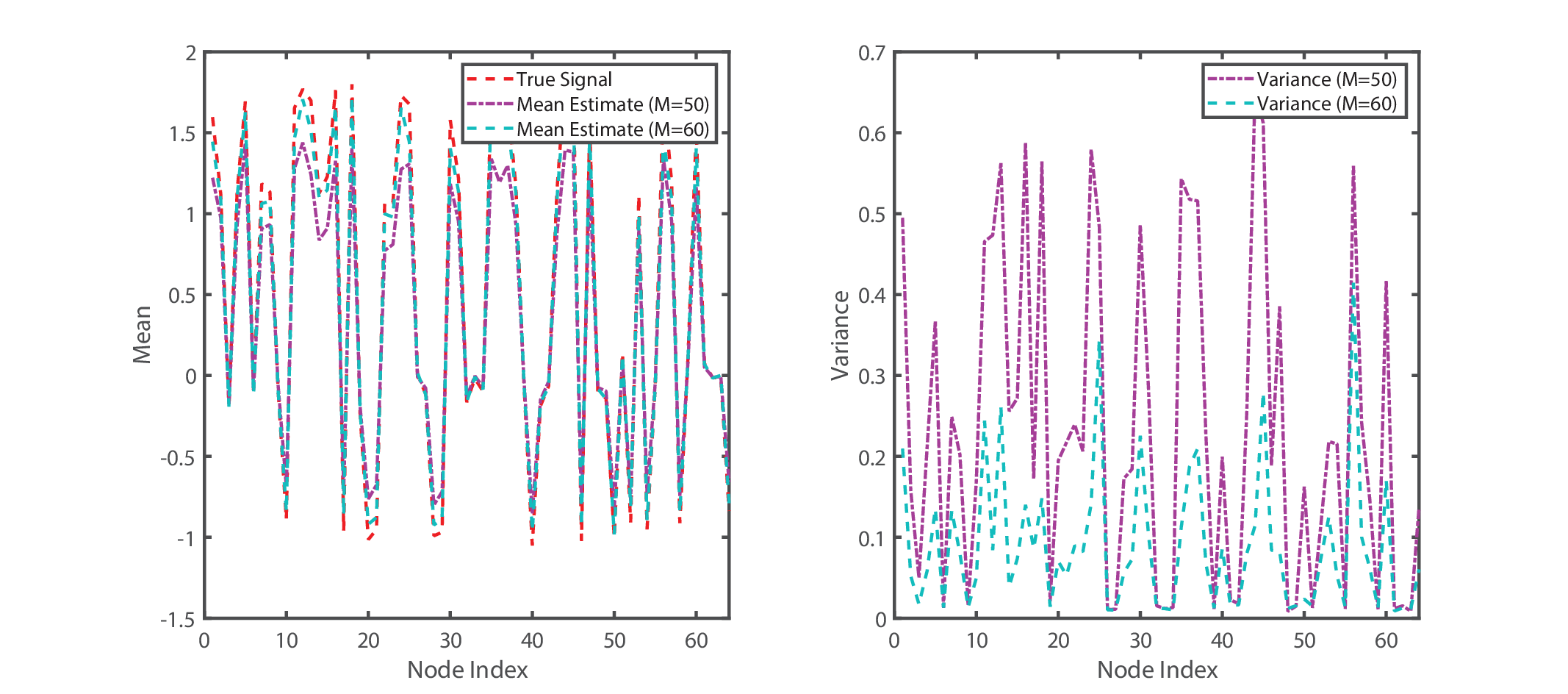}\label{vargmm} } \\[-5ex]
\subfloat[Mean and variance for GMRF signal]{\includegraphics*[width=5.2in]{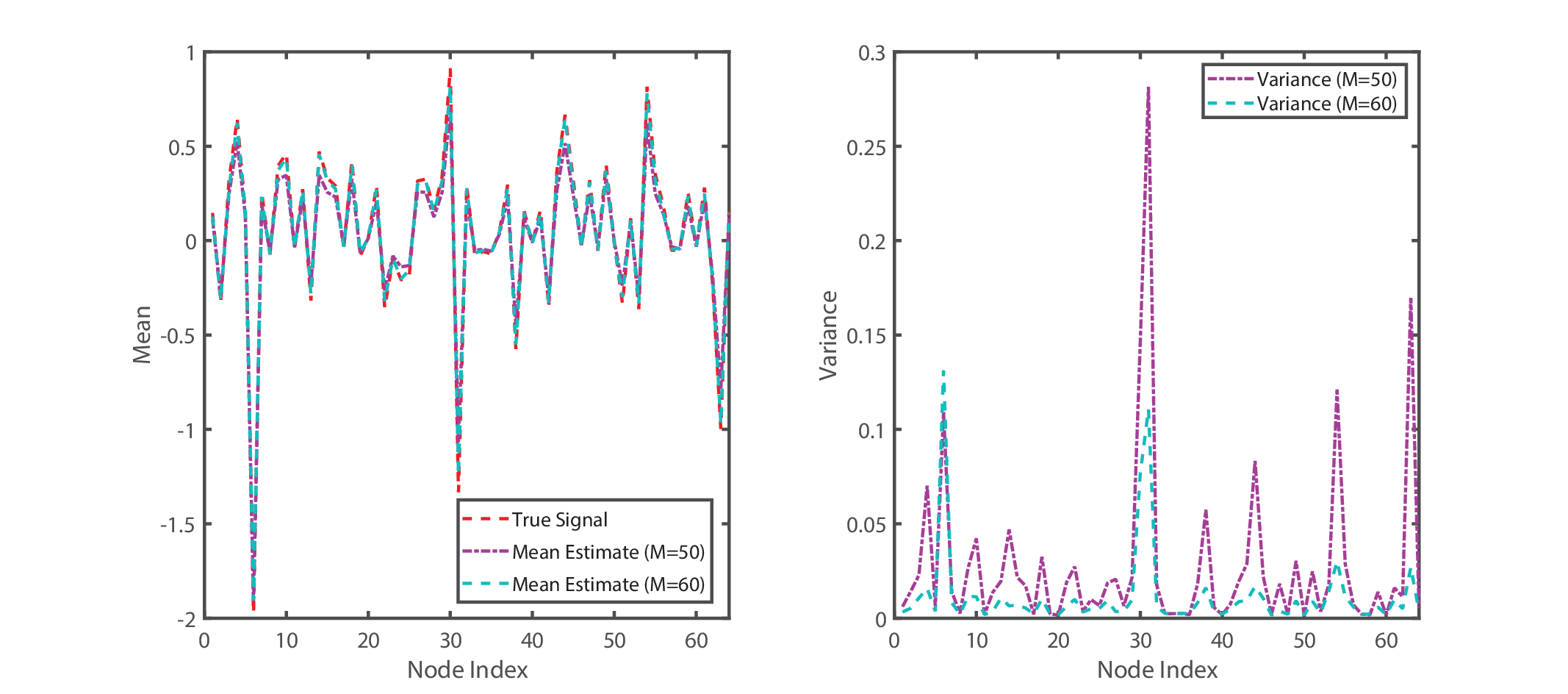}\label{vargmrf} }
\caption{The mean and variance study with 100 iterations when the graph signal is generated from: (a) GMM distribution; and (b) GMRF distribution.}
\label{fig:variance_study}
\end{figure*}

Moreover, we performed a variance study by evaluating learned variances across nodes and components. To perform a variance study with multiple iterations in our proposed BCNN-GSR algorithm, the goal is to statistically evaluate the stability and uncertainty of your recovery algorithm across multiple runs. This includes analyzing the variability of the recovered signal due to random initialization (of filters, weights), noise in observations, and sampling variability (e.g., which nodes are observed). To this end, we run the full experiment 100 times, and at each trial, we generate new training data, noise, and observations. Then, we store all recovered signals and compute signal-wise variance across trials. Figure \ref{fig:variance_study} illustrates the average recovered signal and corresponding variance at each node when the underlying graph signal is generated from GMM and GMRF distributions, respectively. As shown, the mean of the recovered signals over 100 independent trials aligns closely with the true signal in both cases. Furthermore, the proposed algorithm demonstrates low variance in its estimates, indicating consistent and stable recovery performance across trials.

\subsection{Comparison with other GSR algorithms for real-valued data}
\label{simul_temp}
\begin{figure}[!t]
\centering
{\includegraphics[width=.75\linewidth]{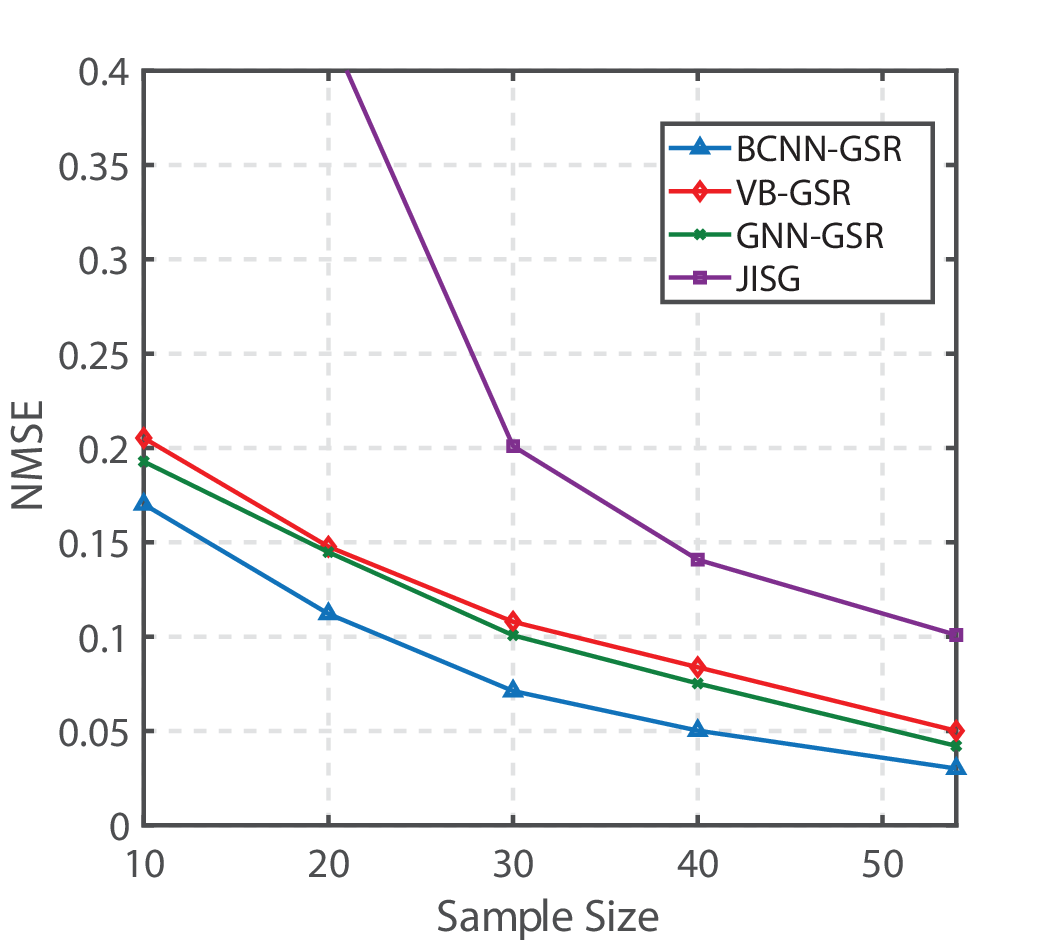}}%
\caption{The NMSE comparison of different algorithms for temperature data.}
\label{temp10dB}
\end{figure}

In this section, the performance of the proposed BCNN-GSR algorithm is tested by the real-valued data. The data is 1-D temperature signals downloaded from the Intel  Berkeley Research lab \cite{Intel04}, which contains $N=54$ sensor nodes and the temperature of sensors are measured as graph signals. Similar to the experiments on synthetic data presented in \ref{simul_GSR}, we compare the performance of our proposed algorithm with the VB-GSR and JISG algorithms. Moreover, for the CNNs-based prior learning algorithm, we consider $K=500$ temperature signals as training data. The i.i.d. Gaussian noise is added to the observation signal to get $\mathrm{SNR}=10dB$. The other settings are the same as previous section. The recovery performance of the competting algorithms is presented in Fig. \ref{temp10dB}.

We can find that the proposed BCNN-GSR algorithm still have the smallest NMSE and achieves the best performance. Note that both BCNN-GSR and VB-GSR algorithms use VB procedure in recovery process, but the proposed BCNN-GSR algorithm outperforms the VB-GSR, which confirms the powerful ability of the proposed CNNs-based prior learning algorithm. In other words, this is due to the fact that the VB-GSR algorithm assumes GMRF model as prior distribution for all types of graph signals, while the proposed BCNN-GSR algorithm learns the prior distribution from the data, and, thus, can suitably estimate the statistical behaviour of the original graph signal.

\section{Conclusion}
\label{sec_concolusion}
In this paper, we investigated the problem of graph signal recovery from noisy and under-sampled observations in a statistical setting, particularly when the graph signal model, i.e., its probability density function (pdf), is unknown. We introduced a novel Bayesian interpretation of convolutional neural networks (CNNs), in which the hidden layers are formulated as Gibbs distributions. This statistical viewpoint enabled us to model the prior distribution of graph signals using a CNN architecture tailored for graphs via Chebyshev polynomial filters.

In particular, we employed a Gaussian Mixture Model (GMM) as the activation function within the CNN, which not only provides a closed-form expression for the prior distribution but also improves denoising performance. The learned prior, combined with a Gaussian likelihood model, allows us to infer the posterior distribution of the unknown signal and noise precision via a variational Bayesian (VB) procedure.

Extensive experiments on both synthetic and real-world datasets demonstrated the superior performance of the proposed BCNN-GSR algorithm compared to several state-of-the-art GSR methods. As future work, we plan to extend our evaluation to larger-scale graph datasets to further assess the scalability and robustness of the proposed method. The use of Chebyshev polynomial graph filters ensures computational efficiency, making the approach well-suited for such extensions.

\end{document}